\documentclass{iaus}
\usepackage{graphicx}
\usepackage{epstopdf}

\newcommand{\msun}{$\rm M_{\odot}$}
\newcommand{\Ha}{$\rm H\alpha$}
\newcommand{\Hb}{$\rm H\beta$}
\newcommand{\lya}{$\rm Ly\alpha$}

\title[Chemical Abundances in High Redshift Galaxies] %% give here short title %%
{Chemical Abundances in Star-Forming Galaxies at High Redshift}

\author[Dawn K. Erb]   %% give here short author list %%
{Dawn K. Erb}
\affiliation{Deparment of Physics\\ University of California Santa Barbara\\ Santa Barbara, CA  93106-9530, USA\\email: {\tt dawn@physics.ucsb.edu}}

\pubyear{2009}
\volume{265}  %% insert here IAU Symposium No.
\pagerange{--}
\jname{Chemical Abundances in the Universe: Connecting First Stars to Planets}
\editors{K. Cunha, M. Spite \& B. Barbuy, eds.}
\begin{document}

\maketitle

\begin{abstract}
A galaxy's metallicity provides a record of star formation, gas accretion, and gas outflow, and is therefore one of the most informative measurements that can be made at high redshift.  It is also one of the most difficult.  I review methods of determining chemical abundances in distant star-forming galaxies, and summarize results for galaxies at $1\lesssim z\lesssim 3$.  I then focus on the mass-metallicity relation, its evolution with redshift, and its uses in constraining inflows and outflows of gas, and conclude with a brief discussion of future prospects for metallicity measurements at high redshift.

\end{abstract}

\firstsection % if your document starts with a section,
\section{Introduction}

A galaxy's metal content is intimately tied to its star formation history and its interactions with the surrounding intergalactic medium (IGM), and thus the measurement of chemical abundances in a galaxy provides, in principle, a detailed fossil record of galaxy evolution.  As succeeding generations of stars are born and die, they enrich the surrounding interstellar medium.  These stars may also drive gas out of the galaxy via powerful supernova-driven winds; such winds carry metals out into the IGM along with the gas.  Galaxies may also accrete new gas from the surrounding IGM; if this gas is low in metallicity, it will decrease the average metallicity of the gas in the galaxy.  Thus an understanding of chemical abundances is a key component of the study of galactic evolution.  

Unfortunately, metallicity is also one of the more difficult properties of galaxies to determine at high redshift, since relatively high signal-to-noise spectra are required and the familiar rest-frame optical emission lines are redshifted into the infrared.   However, enormous progress has been made in this area in recent years, as the advent of sensitive optical and near-IR spectrographs on 8--10 m class telescopes has allowed researchers to obtain rest-frame UV and optical spectra of both emission and absorption lines in galaxies at $1\lesssim z \lesssim 3$.  In this article I review the most widely-used methods of abundance determination in high redshift galaxies and their results, and discuss the mass-metallicity relation at high redshift, its evolution, and its uses in constraining inflows and outflows of gas.

\section{Measuring Galaxy Metallicity}

There are multitudes of methods for measuring metallicity in galaxies, but all require spectra.  At low redshift, the direct or electron temperature ($T_e$) method is preferred; this uses the temperature-sensitive ratio of two transitions of the same ion to determine the electron temperature and hence the metallicity.  Most commonly, the auroral line [O~{\sc iii}]~$\lambda$4363 is used; however, this line is weak under the best conditions and becomes undetectable at metallicities above $\sim0.5$ solar, making its use extremely difficult at high redshifts.  We must therefore turn to other methods.

\subsection{Strong line methods from rest-frame optical emission lines}
Most common among these are the ``strong line" indicators, which are based on the ratios of collisionally excited forbidden lines to hydrogen recombination lines.   Such methods are calibrated either with reference to the $T_e$ method or with photoionization modeling.  There are many such methods, using a variety of different lines (see, e.g., \cite[Kewley \& Ellison 2008]{ke08} for a review and comparison of many of the diagnostics), but perhaps the most widely used for high redshift work to date are the $R_{23}$ and $N2$ methods.

The $R_{23}$ diagnostic (the ratio of the sum of the oxygen lines [O~{\sc ii}]~$\lambda$3727 and [O~{\sc iii}] $\lambda\lambda$4959, 5007 to \Hb; e.g.\ \cite[McGaugh 1991, Zaritsky et al.\ 1994]{m91,zkh94}) is probably the most commonly employed metallicity indicator at low redshifts.  It therefore benefits from being well-studied and used for a wide variety of large samples.  However, the $R_{23}$ method comes with well-known hazards:  it is sensitive to extinction; different calibrations of the ratio can differ in metallicity by as much as 0.5 dex (\cite[Kennicutt et al.\ 2003, Kewley \& Ellison 2008]{kbg03,ke08}); and it is famously double-valued, rising at low metallicities, peaking somewhat below solar metallicity, and falling as the metallicity increases further.  It is therefore not very useful for precise determinations of metallicity near this turnover region; unfortunately, this region is exactly where the metallicities of typical star-forming galaxies at high redshift have been shown to lie.  The wide wavelength separation of the [O~{\sc ii}] and [O~{\sc iii}] lines is a further pitfall for high redshift work, since they fall in different near-IR bands ($J$ and $H$ respectively at $z\sim2$, and $H$ and $K$ at $z\sim3$).   Most near-IR spectrographs are therefore unable to observe them simultaneously, meaning that additional observing time is required to observe in two bands and significant relative calibration uncertainties are introduced.

The $N2$ index, the ratio of [N~{\sc ii}]~$\lambda$6584 to \Ha\ (\cite[Denicol{\'o} et al.\ 2002, Pettini \& Pagel 2004]{dtt02,pp04}), is free from many of the difficulties associated with $R_{23}$.  Because it depends only on two closely spaced lines, it is unaffected by extinction or flux calibration uncertainties, and observations in only one band are required.  On the other hand, its saturation at approximately solar metallicity makes it ineffective for very metal-rich objects, the [N~{\sc ii}]/\Ha\ ratio is sensitive to both the ionization parameter and AGN contamination, and it may be affected by delayed nitrogen production in young, rapidly star-forming galaxies (\cite[P{\'e}rez-Montero \& Contini 2009]{pc09}).  Further difficulties arise because the weak [N~{\sc ii}] line is difficult to detect in the typical low S/N spectra of high redshift objects; because of this, most $N2$ measurements to date have been made either for relatively luminous and metal-rich galaxies or for stacked galaxy spectra.

Clearly both methods suffer from significant uncertainties.  The comparison of the two methods introduces an additional complication.  Thus far, most metallicity measurements at $z\sim2$ have used $N2$, while most of those at $z\sim3$ are based on the oxygen lines.  There is likely to be a significant systematic offset between the two calibrations, and although the best efforts have been made to understand and correct for this offset, it may depend on the physical conditions in the galaxies.  These conditions may be different at high redshift, meaning that a low redshift comparison sample may not be appropriate.  This issue will not be resolved until large samples of high redshift galaxies with metallicities determined with both methods exist; and even then, significant absolute uncertainty will remain, until these indicators can be tied to metallicities determined with the $T_e$ method for the same galaxy sample.  

\subsection{Abundance indicators in the rest-frame UV}

For redshifts above $z\sim2$, the UV spectrum of an object can be obtained much more easily than the optical.  In principle, this allows for exquisitely well-determined abundance measurements, as the rest-frame UV contains resonance lines from a wide variety of elements; with spectra of high resolution and high S/N, the relative abundances of these elements can be determined to fairly high precision.  In practice, spectra of sufficient quality can only be obtained for objects seen in absorption against a bright background source (damped \lya\ systems or GRB host galaxies) or for galaxies greatly boosted in luminosity by gravitational lensing (\cite[Pettini et al.\ 2002]{prs+02}).  

In local starburst galaxies, the UV low ionization interstellar absorption lines have been shown to correlate with metallicity, as has the blended stellar and interstellar high ionization line {\sc C iv}~$\lambda$1550 (\cite[Heckman et al.\ 1998]{hrl+98}).  These correlations are potentially very useful for determining metallicity, as these lines are among the strongest features in the UV spectra of high redshift starbursts, and indeed, both sets of features have been used to estimate the metallicities of distant galaxies (\cite[Mehlert et al.\ 2002, Ando et al.\ 2007]{aoi+07,mna+02}).  However, the relationship between the strength of these lines and the metallicity of the absorbing gas remains uncalibrated at high redshift; detailed spectral studies indicate that the interstellar line strengths depend primarily on the velocity dispersion and covering fraction of outflowing gas, with metallicity at best a secondary effect because the lines are saturated (\cite[Shapley et al.\ 2003]{ssp+03}).  Until an improved calibration is done, such metallicity estimates must therefore be treated skeptically.

In an effort to put UV-based metallicity indicators on a more systematic footing, \cite{rpl+04} have recently used spectral synthesis and non-LTE model atmospheric codes to model the dependence of stellar photospheric features on metallicity.  They suggest the use of several broad but fairly weak features as abundance indicators, with the most promising being a complex of Fe~{\sc iii} transitions between 1935 and 2020 \AA.  While the weakness of these features has proven an obstacle to their use in the typical low S/N spectra of high redshift galaxies, they have been used with some success on composite spectra (\cite[Erb et al.\ 2006, Halliday et al.\ 2008]{esp+06,hdc+08}).  However, recent attempts to use these indicators on spectra of lensed galaxies have proven problematic, as features in the galaxy spectra are not always reproduced well by the models (\cite[Quider et al.\ 2009, 2010]{qpss09}).  

In summary, the rest-frame UV is rich in metallicity-dependent features.  However, due to the complex nature of many of these lines and the insufficient spectral resolution and S/N of much of the data, this is a regime for which widely applicable metallicity indicators have yet to be developed.

\section{The Metallicities of Star-Forming Galaxies at \boldmath$1\lesssim z\lesssim3$\unboldmath}

It has been about ten years since the first measurements of metallicity in star-forming galaxies at high redshift.  From these early results (\cite[Kobulnicky \& Koo 2000, Pettini et al.\ 2001]{kk00,pss+01}), two things immediately became clear: luminous high redshift galaxies are significantly enriched, with metallicities well above the very low values seen in many DLAs at this epoch, and, at a given metallicity, high redshift galaxies are significantly more luminous than their local counterparts.  Further work showed that massive ($M_{\star}\gtrsim10^{11}$ \msun) galaxies have approximately solar metallicities, even at $z\sim2$, when the universe was only $\sim20$\% of its present age (\cite[Shapley et al.\ 2004, van Dokkum et al.\ 2004, F{\"o}rster Schreiber et al.\ 2006]{sep+04,vff+04,fgl+06}). 

Because of their significantly boosted luminosity, gravitationally lensed galaxies have been the subjects of the most detailed high redshift abundance measurements to date.  First and still the best-studied among these is the $z=2.7$ galaxy MS 1512-cB58, a typical Lyman break galaxy magnified by a factor of $\sim30$ by a foreground cluster.  Initial optical (\cite[Teplitz et al.\ 2000]{tmb+00}) and UV (\cite[Pettini et al.\ 2000]{psa+00})  measurements indicated a metallicity of $\sim1/4$--1/3 Z$_{\odot}$; more interestingly, however, the brightness of the galaxy allows for high resolution spectra from which relative abundances of different elements can be measured.  These measurements refine the abundance of oxygen and other $\alpha$ elements to $\sim2/5$ solar, while nitrogen and the iron peak elements are underabundant by a factor of $\sim3$.  These values are consistent with the rapid star formation and young age of cB58, as such a pattern would be expected if most of the metal enrichment had occurred within the last $\sim300$ Myr, on the timescale for the release of nitrogen from intermediate mass stars (\cite[Pettini et al.\ 2002]{prs+02}).  High resolution UV spectra have now been obtained for other lensed galaxies as well, but this kind of elemental abundance analysis has proven difficult because the \lya\ profile does not always allow for a measurement of the hydrogen column density, and intervening absorption systems may make other lines difficult to measure  (\cite[Quider et al.\ 2009, 2010]{qpss09}).  

Gravitational lensing also allows the detection of galaxies that would otherwise be too faint to observe at all.  \cite{yk09} have recently reported the first detection of the [O {\sc iii}] $\lambda$ 4363 line at high redshift, in a lensed low mass ($M_{\star}=4.4 \times 10^8$ \msun) galaxy at $z=1.7$.  The electron temperature metallicity derived from this measurement is quite low, 12 + log (O/H) = 7.5; this is 0.6 dex lower than the metallicity derived from the $R_{23}$ method for the same galaxy, a discrepancy which is perhaps worrying; on the other hand, offsets of this magnitude between metallicity diagnostics, or even between different calibrations of the same diagnostic, are not uncommon, and $R_{23}$-derived metallicities may be high compared to many other indicators (\cite[Kewley \& Ellison 2008]{ke08}).

\section{The Mass-Metallicity Relation}

The correlation between galaxy mass and metallicity has been well-known for some time in the local universe (\cite[Lequeux et al.\ 1979, Tremonti et al.\ 2004]{lpr+79,thk+04}).  The shape and amplitude of this relationship depends on all of the key processes in galaxy evolution: the conversion of gas to stars and the subsequent enrichment of the gas by stellar winds and supernovae; the loss of gas and metals via galactic outflows; and the accretion of new, metal-poor gas, which dilutes the metallicity of the gas in the galaxy.  Measurement of the mass-metallicity relation therefore has the potential to constrain these fundamental processes.

Because large spectroscopic samples of the rest-frame optical emission lines are needed in order to determine chemical abundances, the mass-metallicity relation has only recently been measured at $z>1$.  \cite{esp+06} used composite \Ha\ and [{\sc N~ii}] spectra of 87 galaxies divided into six bins by stellar mass to measure the mass-metallicity relation at $z\sim2$, finding that at a given stellar mass, $z\sim2$ galaxies are $\sim0.3$ dex lower in metallicity than star-forming galaxies in the local universe.  Many additional measurements have now been made at $1\lesssim z\lesssim 3$ (\cite[P{\'e}rez-Montero et al.\ 2009, Liu et al.\ 
2008, Hayashi et al.\ 2009, Law et al.\  2009, Maiolino et al.\ 2008, Mannucci et al.\ 2009]{pcl+09, lsc+08, hms+09, lse+09, mng+08, mcm+09}); these studies have used spectra of individual galaxies rather than composites, an important step toward understanding the scatter in the relationship at high redshift.

\subsection{Redshift Evolution of the Mass-Metallicity Relation}

Measurements of the mass-metallicity relation at high redshift consistently show that at a given stellar mass, distant star-forming galaxies are lower in metallicity than those in the local universe; the amount of this offset generally increases with increasing redshift.  At $z\sim0.7$, \cite{sgl+05} find that galaxies are $\sim0.1$--0.2 dex lower in metallicity at a given mass than in the local universe, with considerable scatter and a larger offset at lower masses.  At $z\sim1$, \cite{pcl+09} find an offset of $\sim0.3$ dex, comparable to that found at $z\sim2$ by \cite{esp+06}, while at $z\sim3.5$, \cite{mng+08} and \cite{mcm+09} find that galaxies are $\sim6$ times lower in metallicity than in the local universe.

There are a variety of reasons to expect that galaxies at higher redshifts may be less enriched:  they are on average younger than galaxies in the local universe and probably have higher gas fractions (\cite[Erb et al.\ 2006b]{ess+06mass}); they may be accreting significant amounts of metal-poor gas (e.g.\ \cite[Kere{\v s} et al.\ 2005]{kkwd05}); and they drive powerful outflows, expelling gas and metals into the IGM (\cite[Pettini et al.\ 2000, 2001]{psa+00,pss+01}).  Nevertheless, the comparison of metallicities across a wide redshift range is difficult, and there are several issues which must be addressed before the redshift evolution of the mass-metallicity relation can be reliably quantified.  First, it is usually not possible to observe the same set of emission lines for all of the galaxy samples, meaning that a variety of different metallicity indicators have been used for the above studies.  While the best efforts have been made to adjust these calibrations to be consistent with each other, the conversions between metallicity indicators are necessarily based on local galaxy samples (e.g.\ \cite[Kewley \& Ellison 2008]{ke08}), and physical conditions affecting the ratios of the strong emission lines may evolve with redshift.

Second, it should also be realized that this shift in metallicity with redshift does not represent an evolutionary sequence:  the higher redshift samples are probably not the progenitors of the galaxies at lower redshifts (\cite[Erb et al.\ 2006a, Maiolino et al.\ 2008, Conroy et al.\ 2008]{esp+06,mng+08,cst+08}).  Rather, it represents the average metallicity of luminous star-forming galaxies at a given epoch.  A further issue related to the above two points is the apparently large evolution between $z\sim2.3$ and $z\sim3.5$.  Over the $\sim11$ Gyr between $z\sim0$ and $z\sim2$, the average metallicities decrease by a factor of $\sim2$--2.5; the same decrease is seen over the $\sim1$ Gyr between $z\sim2.3$ and $z\sim3.5$.  Models and simulations currently fail to explain this rapid evolution (\cite[Maiolino et al.\ 2008]{mng+08}).  Finally, several studies of the high redshift mass-metallicity relation have noted that the offset with respect to local galaxies appears to be larger at low masses; in other words, massive galaxies have metallicities closer to those of massive galaxies in the local universe.  Such a pattern would not be surprising if the typical mass of galaxies hosting the bulk of star formation shifts to lower masses at lower redshifts (e.g.\ \cite[Bundy et al.\ 2006]{bec+06}).

\section{Constraining Gas Flows with Metallicity}

Because the mass-metallicity relation depends on inflows and outflows of gas as well as on star formation, it is a potentially useful tool to assess the importance of these processes in galaxies.  If a galaxy is a closed box with no inflows or outflows of gas, the metallicity $Z$ and the gas fraction $\mu$ are simply related:
\begin{equation}
Z=y \ln(1/\mu),
\label{eq:closedbox}
\end{equation}
where $y$ is the yield, the ratio of the mass of metals produced and ejected by stars to the mass locked in long-lived stars and remnants (\cite[Edmunds 1990]{e90}).  This relation can be inverted to define the effective yield $y_{\rm eff}$:
\begin{equation}
y_{\rm eff}=\frac{Z}{\ln(1/\mu)}.
\end{equation}
If the galaxy indeed behaves as a closed box, the effective yield will be equal to the true yield, and either the expulsion of gas and metals from the galaxy or the accretion of metal-poor gas lower the effective yield (\cite[Edmunds 1990]{e90}).  Thus, measurements of metallicity and gas fraction can reveal the presence of inflows or outflows of gas.  In the local universe, $y_{\rm eff}$ is observed to be lower in lower mass galaxies; this is interpreted as a signature of preferential loss of gas and metals from the shallower potential wells of lower mass galaxies  (\cite[Tremonti et al.\ 2004]{thk+04}; see also \cite[Dalcanton 2007]{d07}).

At high redshifts, current technology does not yet allow direct measurement of the gas fractions of typical star-forming galaxies.  Gas masses and gas fractions are therefore usually estimated from the star formation rate surface density, assuming the local Kennicutt-Schmidt law (\cite[Kennicutt 1998]{k98schmidt}).  Using gas fractions determined in this way, effective yields have been calculated for galaxies at $z\sim2$ (\cite[Erb et al.\ 2006a]{esp+06}) and at $z\sim3.5$ (\cite[Maiolino et al.\ 2008, Mannucci et al.\ 2009]{mng+08, mcm+09}).  In both cases, the effective yield is seen to \textit{increase} at lower masses, the opposite of the trend seen in the local universe. 

Further estimates of the relative inflow and outflow rates can be obtained from the Kennicutt-Schmidt law: it can be shown that if the local K-S law holds at high redshift, the extended star formation histories observed in galaxies require the accretion of new gas at approximately the rate it is lost to outflows and star formation (\cite[Erb 2008]{e08}).  Equation \ref{eq:closedbox} can then be modified to include both inflows and outflows of gas (parameterized as a fixed fraction of the star formation rate for simplicity).  Incorporating the gas inflow rate required by the K-S law then allows a determination of the outflow rate.  The metallicities and gas fractions of the $z\sim2$ galaxies are best fit with an outflow rate approximately equal to the star formation rate, and an inflow rate roughly equal to the combined outflow and star formation rates (\cite[Erb 2008]{e08}).  This model is also a good fit to the metallicities and gas fractions at $z\sim3.5$ (\cite[Mannucci et al.\ 2009]{mcm+09}).

While these values are in general agreement with outflow rates estimated from observations (\cite[Pettini et al.\ 2000]{psa+00}) and theoretically predicted inflow rates (e.g.\  \cite[Kere{\v s} et al.\ 2005]{kkwd05}), significant uncertainties remain.  In addition to the indirectly inferred gas fractions and the systematic uncertainties associated with the metallicity measurements, these models require assumptions of the metallicity of the inflowing and outflowing gas, neither of which can currently be measured at high redshift.  The appropriate value of the true yield $y$ is also not well known.  In short, measurements of metallicity and gas fraction at high redshift provide a model for gas inflows and outflows consistent with observational data and theoretical expectations, but because of the large number of free parameters this model is not a unique solution.  It does seem clear, however, that the canonical low redshift model of increasingly efficient removal of metals in low mass galaxies does not match the observed data at high redshift  (\cite[Erb et al.\ 2006a, Mannucci et al.\ 2009]{esp+06,mcm+09}).  Additional observations of lower mass galaxies are needed to confirm this result.

\section{Summary and Future Outlook}

The measurement of chemical enrichment in galaxies at high redshift is a difficult problem, since faint object spectroscopy is always required and large systematic uncertainties are associated with all metallicity diagnostics.  Nevertheless, there is an increasingly large sample of measurements at $1 \lesssim z \lesssim 3$.  Galaxies at these redshifts are significantly enriched, with typical oxygen abundances of a few tenths of solar; massive ($M_{\star}\gtrsim 10^{11}$ \msun) galaxies appear to have reached approximately solar metallicities by $z\sim2$.  The mass-metallicity relation is already in place at $z\sim3.5$, and shows significant evolution with redshift, in the sense that at a given stellar mass, high redshift galaxies are lower in metallicity.   Unlike at $z=0$, the mass-metallicity relation cannot be explained by a simple model in which outflows are more effective at removing metals from galaxies at lower masses.  Rather, the high redshift mass-metallicity relation is best fit with combined inflow and outflow models, where the outflow rate is approximately equal to the star formation rate and significant gas accretion is required.  While this picture is in general agreement with the (very limited) observational data and theoretical expectations regarding inflows and outflows of gas at high redshifts, many uncertainties and free parameters remain.

The number of $1 \lesssim z \lesssim 3$ galaxies with metallicity measurements will grow by leaps and bounds in the very near future, as multi-object near-IR spectrographs are put into place on 8--10 m telescopes.  These will allow larger samples of larger numbers of emission lines, and deep, higher S/N spectra and spectra of fainter objects as it becomes more worthwhile to spend large amounts of time on a single observation.   Less optimistically, measurements beyond $z\sim3$ will remain extremely difficult, as the strong [O {\sc iii}] lines and H$\beta$ shift beyond the $K$-band and become inaccessible from the ground; a space-based near-IR spectrograph such as that planned for JWST will be needed for robust measurements of metallicities in these galaxies.  The other option is to turn to the rest-frame UV, which remains accessible; however, although there are many metallicity-dependent features here, their use as diagnostics has proven difficult due to the complicated origin of the spectral features and the low S/N and resolution of the spectra.  As these data are relatively easy to obtain and will dominate very high redshift observational efforts for some time to come, improved calibrations of the metallicity dependence of the strongest features in the rest-frame UV may offer the best hope for constraining the abundances of the most distant galaxies.  Metallicity measurements from new, larger samples of rest-frame optical emission lines at $z\sim2$--3 will provide the necessary data for these calibrations.

\end{document}